\documentclass[preprint,showpacs,amsmath]{revtex4}

\usepackage{times}
\usepackage{hyperref}

\usepackage{graphicx}
\usepackage{dcolumn}
\usepackage{bm}

\begin{document}

\title{
$V_{us}$ from hyperon semileptonic decays
}

\author{
Rub\'en Flores-Mendieta
}

\affiliation{
Instituto de F{\'\i}sica, Universidad Aut\'onoma de San Luis Potos{\'\i}, \'Alvaro Obreg\'on 64, Zona Centro, San Luis
Potos{\'\i}, S.L.P.\ 78000, M\'exico
}

\date{\today}

\begin{abstract}
A model-independent determination of the CKM matrix element $V_{us}$ from five measured strangeness-changing
hyperon semileptonic decays is performed. Flavor $SU(3)$ symmetry breaking effects in the leading vector and
axial-vector form factors are analyzed in the framework of the $1/N_c$ expansion of QCD. A fit to experimental data
allows one to extract the value $V_{us}=0.2199\pm 0.0026$, which is comparable to the one from $K_{e3}$ decays.
This reconciliation is achieved through second-order symmetry breaking effects of a few percent in
the form factors $f_1$, which increase their magnitudes over their $SU(3)$ predictions.

\end{abstract}


\maketitle

\section{Introduction}

Hyperon semileptonic decays (HSD) play a decisive role in our understanding of the interplay between weak and strong
interactions and the Cabibbo-Kobayashi-Maskawa (CKM) quark-mixing matrix.
At present, the determinations of $V_{ud}$ and $V_{us}$ provide the most precise constraints on the size
of the CKM matrix elements.
It has been argued that $K_{e3}$ decays offer possibly the cleanest way to extract a precise value of $V_{us}$ rather
than HSD. From the theoretical point of view, the leptonic part
of both semileptonic processes is unambiguous. In contrast, the hadronic part is deeply affected by flavor $SU(3)$ symmetry
breaking in the form factors. For $K_{e3}$ decays, this is a minor problem because only the vector part of the weak current has a
nonvanishing contribution and only two form factors appear. In addition, such form factors are protected by the Ademollo-Gatto theorem
\cite{ag} against $SU(3)$ breaking corrections to lowest order in $(m_s-\hat{m})$ so that the theoretical approach to compute
them is under reasonable control within the limits of experimental precision. On the contrary, HSD are
considerably more complicated than $K_{e3}$ decays due to the participation of vector and axial-vector currents, which leads to the
appearance of many more form factors. Although the leading vector form factors are also protected by the Ademollo-Gatto theorem,
the analysis
of HSD data has larger theoretical uncertainties because of first-order $SU(3)$ breaking effects in the axial-vector form
factors.

Indeed, the current value of $V_{us}$ recommended by the Particle Data Group \cite{part} is the one from $K_{e3}$ decays, namely,
\begin{equation}
V_{us} = 0.2200 \pm 0.0026. \label{eq:vuskl3}
\end{equation}
Recent studies of $K_{e3}$ decays \cite{ktev,sher,batt}, HSD \cite{cabb}, and lattice gauge theory \cite{mar} suggest larger
values of $V_{us}$, in disagreement with early determinations \cite{leut}. This discrepancy is an outstanding problem
and should be addressed.

Inspired by those facts, in this paper we perform a detailed model-independent analysis of the determination of $V_{us}$ from five already
observed $|\Delta S|=1$ HSD. The goals in performing this study are to confirm the value of $V_{us}$ obtained from $K_{e3}$ decays,
Eq.~(\ref{eq:vuskl3}), and to use the form factors to achieve a better understanding of the hadronic structure.

In order to have a precise and reliably determination of $V_{us}$, we systematically consider two major approaches. First,
we incorporate radiative corrections to various measurable quantities relevant for experimental analyses and include
the momentum-transfer contributions
of the form factors. And second, we analyze $SU(3)$ symmetry breaking effects into the leading vector and axial-vector
form factors in the framework of the $1/N_c$ expansion of QCD, following the lines of Ref.~\cite{rfm98}.
The resultant theoretical expressions are thus compared with the available experimental data on HSD \cite{part}, allowing
an extraction of $V_{us}$. Here we need to point out a slight difference between our procedure and the one of Ref.~\cite{rfm98}.
There, a global fit of HSD and pionic decays of the decuplet baryons was performed,
whereas in our case we concentrate only on the $|\Delta S|=1$ sector.

This work is organized as follows. In Sec.~\ref{sec:hsd} we provide some theoretical issues on HSD. In Sec.~\ref{sec:ncexp}
we give a general overview of the $1/N_c$ expansion of baryon operators whose matrix elements yield the HSD form factors.
In Sec.~\ref{sec:rates}-\ref{sec:mod} we perform detailed
comparisons of the theoretical expressions with the current experimental data on HSD \cite{part}
through several fits under various assumptions. We present results and
conclusions in Sec.~\ref{sec:concl}. In Appendix \ref{sec:appa} we provide numerical formulas for the integrated observables
used in our analysis.

\section{\label{sec:hsd}Hyperon semileptonic decays}

In this section we will review our notation and conventions. For definiteness, let us consider the hyperon semileptonic decay
\begin{equation}
B_1 \to B_2 + \ell + \overline{\nu}_l, \label{eq:ec1}
\end{equation}
where $B_1$ and $B_2$ are spin-1/2 hyperons, $\ell$ is the charged lepton ($\ell=e,\nu$), and
$\overline{\nu}_\ell$ is the accompanying antineutrino or neutrino, as the case may
be. The four-momenta and masses of the particles involved in process (\ref{eq:ec1}) are denoted hereafter by
$p_1=(E_1,{\mathbf p}_1)$ and $M_1$, $p_2=(E_2,{\mathbf p}_2)$ and $M_2$, $l=(E,{\mathbf l})$ and $m$, and
$p_\nu=(E_\nu^0,{\mathbf p}_\nu)$ and $m_\nu$, respectively.

The low-energy weak interaction Hamiltonian for semileptonic processes reads
\begin{eqnarray}
H_W = \frac{G}{\sqrt{2}} J_\alpha L^\alpha + \mathrm{H.c.}, \label{eq:ec2}
\end{eqnarray}
where $L_\alpha$ and $J_\alpha$ denote the leptonic and hadronic currents, respectively. The former is given by
\begin{eqnarray}
L^\alpha = \overline \psi_e \gamma^\alpha (1 - \gamma_5) \psi_{\nu_e} + \overline \psi_\mu \gamma^\alpha (1 - \gamma_5)
\psi_{\nu_\mu}, \label{eq:ec3}
\end{eqnarray}
whereas $J_\alpha$, expressed in terms of the vector $(V_\alpha)$ and axial-vector $(A_\alpha)$ currents, can be written as
\begin{subequations}
\label{eq:ec4}
\begin{eqnarray}
J_\alpha & = & V_\alpha - A_\alpha, \\
V_\alpha & = & V_{ud} \overline u \gamma_\alpha d + V_{us} \overline u \gamma_\alpha s, \\
A_\alpha & = & V_{ud} \overline u \gamma_\alpha \gamma_5 d + V_{us} \overline u \gamma_\alpha \gamma_5 s.
\end{eqnarray}
\end{subequations}
Here $G$ is the weak coupling constant, and $V_{ud}$ and $V_{us}$ are the appropriate elements of the CKM matrix.

The matrix elements of $J_\alpha$ between spin-$1/2$ states can be written as
\begin{eqnarray}
\langle B_2 |V_\alpha|B_1 \rangle & = & V_{\rm CKM} \, \overline u_{B_2} (p_2) \left[ f_1(q^2) \gamma_\alpha + \frac{f_2(q^2)}{M_1}
\sigma_{\alpha\beta}q^\beta + \frac{f_3(q^2)}{M_1} q_\alpha \right] u_{B_1}(p_1), \label{eq:ec5}
\end{eqnarray}
\begin{eqnarray}
\langle B_2 |A_\alpha|B_1 \rangle & = & V_{\rm CKM} \, \overline u_{B_2} (p_2) \left[g_1(q^2) \gamma_\alpha + \frac{g_2(q^2)}{M_1}
\sigma_{\alpha\beta}q^\beta + \frac{g_3(q^2)}{M_1} q_\alpha \right] \gamma_5 u_{B_1}(p_1), \label{eq:ec6}
\end{eqnarray}
where $q \equiv p_1 - p_2$ is the four-momentum transfer, $u_{B_1}$ and $\overline{u}_{B_2}$ are the Dirac spinors of the
corresponding hyperons, and $V_{\rm CKM}$ is either $V_{ud}$ or $V_{us}$. In this work we adopt the metric and
$\gamma$-matrix conventions of Ref.~\cite{gk}. The quantities $f_1(q^2)$ and $g_1(q^2)$ are the vector and axial-vector form
factors, $f_2(q^2)$ and $g_2(q^2)$ are the weak magnetism and electricity form factors, and $f_3(q^2)$ and $g_3(q^2)$ are the
induced scalar and pseudoscalar form factors, respectively. Time reversal invariance requires the form factors to be real.
$f_3(q^2)$ and $g_3(q^2)$, for electron or positron emission, have negligible contributions to the decay rate due to the
smallness of the factor $(m/M_1)^2$ which comes along with them. Therefore, to a high degree of accuracy, the $e$-modes of
HSD are described in terms of four, rather than six, form factors. In contrast, for $\mu$-modes although the factor
$(m/M_1)^2$ is still small, $f_3(q^2)$ and $g_3(q^2)$ may contribute with some significance and should be retained. For
convenience, here we introduce the definitions $f_i \equiv f_i(0)$ and $g_i \equiv g_i(0)$, with $i = 1,2,3$.

\subsection{Differential decay rate}

The transition amplitude for process (\ref{eq:ec1}) can be constructed from the product of the matrix
elements of the hadronic and leptonic currents \cite{gk}. From this amplitude, the differential decay rate of HSD,
denoted here by $d\Gamma$, can be derived by using standard techniques \cite{gk,rfm97}. For the three-body decay
(\ref{eq:ec1}) different choices of the five
relevant variables in the final states will lead to appropriate expressions for $d\Gamma$. In Ref.~\cite{gk}, for instance,
detailed expressions have been obtained for $d\Gamma$ in the rest frame of $B_1$ $[B_2]$ when such hyperon
is polarized along the direction
$s_1$ $[s_2]$, and with the charged lepton $\ell$ and neutrino going into the solid angles $d\Omega_\ell$ and $d\Omega_\nu$,
respectively. Similarly, in Refs.~\cite{rfm97,rfm01} $d\Gamma$ has been obtained, in the rest frame of $B_1$,
by leaving the electron and emitted hyperon
energies as the relevant variables along with some suitable angular variables.

In all the above cases the differential decay rate can be written, in the most general case, as
\begin{equation}
d\Gamma = G^2 d\Phi_3 \left[A_0^\prime - A_0^{\prime \prime} \, {\hat {\mathbf s}} \cdot {\hat {\mathbf p}} \right],
\end{equation}
where $d\Phi_3$ is an element of the appropriate three-body phase space and $A_0^\prime$ and $A_0^{\prime \prime}$ depend on the
kinematical variables and are quadratic functions of the form factors. The scalar product ${\hat {\mathbf s}} \cdot
{\hat {\mathbf p}}$, where ${\hat {\mathbf s}}$ denotes the spin of either $B_1$ or $B_2$ and ${\hat {\mathbf p}}=
{\hat {\mathbf l}}, {\hat {\mathbf p}_2}, {\hat {\mathbf p}_\nu}$, represents the angular correlation between such spin and
the three-momentum of the corresponding particle \cite{rfm97,rfm01}.

\subsection{Integrated observables}

When experiments in HSD have low statistics one cannot perform a detailed analysis of the differential decay
rate $d\Gamma$. One is thus led to produce some integrated observables instead, namely, the total decay rate $R$ and angular
correlation and asymmetry coefficients. The definitions of these observables entail only kinematics and do not assume any
particular theoretical approach. For example, the charged lepton-neutrino angular correlation coefficient is defined as
\begin{equation}
\alpha_{\ell\nu} = 2 \frac{N(\Theta_{\ell\nu} < \pi/2) - N(\Theta_{\ell\nu} > \pi/2)}{N(\Theta_{\ell\nu} < \pi/2)
+ N(\Theta_{\ell\nu}> \pi/2)},
\end{equation}
where $N(\Theta_{\ell\nu} < \pi/2)$ $[N(\Theta_{\ell\nu} > \pi/2)]$ is the number of charged lepton-neutrino pairs emitted in
directions that make an angle between them smaller [greater] than $\pi/2$. Similar expressions can be derived for the charged
lepton $\alpha_\ell$, neutrino $\alpha_\nu$, and emitted hyperon $\alpha_B$ asymmetry coefficients, this time $\Theta_\ell$,
$\Theta_\nu$, and $\Theta_B$ being the angles between the $\ell$, $\nu$, and $B_2$ directions and the polarization of
$B_1$, respectively. When the polarization of the emitted hyperon is observed, two more asymmetry coefficients, $A$ and $B$,
can be defined \cite{gk}. If the charged lepton mass can be neglected it is rather straightforward to compute approximate
theoretical
expressions for these observables. All this has been done in Ref.~\cite{gk} for a number of decays. For the
uncorrected total decay rate one has
\begin{eqnarray}
R^0 & = & G^2 \frac{(\Delta M)^5}{60\pi^3} \left[\left(1-\frac32 \beta + \frac67 \beta^2\right)f_1^2 + \frac47 \beta^2 f_2^2
+ \left(3 - \frac92 \beta + \frac{12}{7} \beta^2 \right) g_1^2 \right. \nonumber  \\
&  & \mbox{} + \left. \frac{12}{7} \beta^2 g_2^2 + \frac67 \beta^2 f_1f_2 + (-4\beta+6\beta^2)g_1g_2 \right], \label{eq:gam}
\end{eqnarray}
where $\beta = (M_1-M_2)/M_1$ and the superscript 0 on a given observable is used as an indicator that no radiative
corrections have been incorporated into it.
In Eq.~(\ref{eq:gam}), although the form factors have been assumed to be constant, their
$q^2$-dependence cannot always be neglected since they can give a noticeable contribution. In order to obtain expressions
correct to order ${\mathcal O}(q^2)$, the $q^2$-dependence of $f_2$ and $g_2$ can be ignored because they already contribute
to order ${\mathcal O}(q)$ to the decay rate. For $f_1(q^2)$ and $g_1(q^2)$, however, a linear expansion in $q^2$ is enough
because higher powers amount to negligible contributions to the decay rate, no larger than a fraction of a percent. Thus,
\begin{equation}
f_1(q^2) = f_1(0) + \frac{q^2}{M_1^2} \lambda_1^f, \qquad \qquad g_1(q^2) = g_1(0) + \frac{q^2}{M_1^2} \lambda_1^g,
\end{equation}
where the slope parameters $\lambda_1^f$ and $\lambda_1^g$ are both of order unity \cite{gk}. A dipole parametrization
for the leading form factors such as $f(q^2) = f(0)/(1-q^2/M^2)^2$ yields
\begin{equation}
\lambda_1^f = \frac{2M_1^2 f_1}{M_V^2}, \qquad \qquad \lambda_1^g = \frac{2M_1^2 g_1}{M_A^2}, \label{eq:slop}
\end{equation}
where $M_V= 0.97$ GeV and $M_A=1.11$ GeV for $|\Delta S|=1$ HSD \cite{gk}.

For more precise formulas and when the charged lepton mass is retained, one needs to numerically integrate over the
kinematical variables the expressions for $d\Gamma$ and angular coefficients already given in previous works
\cite{gk,rfm97,rfm01}. Concerning this, Ref.~\cite{gk} provides complete numerical formulas for the decay rates and angular
coefficients of the 16 $e$-mode and 10 $\mu$-mode HSD. These formulas, however, are almost 20 years old and the current
experimental data on hyperon masses \cite{part} introduce modifications to them which need to be accounted for. We have
recalculated and
updated the formulas for the uncorrected integrated observables of five HSD we are concerned with in the present analysis. They are
listed in Appendix A for the sake of completeness.

\subsection{Radiative corrections}

Experiments on HSD have gradually become sensitive enough to require radiative corrections to the integrated observables.
However, the calculation of radiative corrections to processes involving hadrons has been a long standing problem. Despite
the outstanding progress achieved in the understanding of the fundamental interactions with the Standard Model \cite{part}, no first
principle calculation of radiative corrections is yet possible. These corrections thus become committed to model dependence
and experimental analyses that use them also become model dependent. Even if the model dependence arising from the virtual radiative
corrections cannot be eliminated, an analysis in neutron beta decay further extended to HSD \cite{sirlin} shows that
to orders $(\alpha/\pi)(q/M_1)^0$ and $(\alpha/\pi)(q/M_1)$ such model-dependence amounts to some constants, which can be
absorbed into the form factors originally defined in the matrix elements of the hadronic current. In addition, the theorem of
Low in its version by Chew \cite{low} can be used to show that to these two orders of approximation the bremsstrahlung
radiative corrections depend only on both the non-radiative form factors and the static electromagnetic multipoles of the
particles involved so that no model-dependence appears in this other part of the radiative corrections. Within these
orders of approximation one is left with general expressions which can be used in
model-independent analyses \cite{gk,rfm97,rfm01}.

The radiative corrections to order $(\alpha/\pi)(q/M_1)^0$ to all the integrated observables of HSD
referred to above have been computed in Ref.~\cite{gk}.
There it was shown that to this order of approximation, the angular and asymmetry coefficients for both
$e$- and $\mu$-mode do not get affected by these corrections, so to a good approximation, $\alpha \simeq \alpha^0$, where
$\alpha$ stands for any of the angular coefficients considered here. In contrast, the total decay rate $R$ is corrected as
$R = R^0 [1 + (\alpha/\pi) \Phi]$, where $R^0$ is the uncorrected decay rate and $\Phi$ comes from the model-independent
part of radiative corrections. The function $\Phi$ can be obtained from Eqs.~(5.25) and (5.28) of Ref.~\cite{gk}; their
numerical values for several decays are listed in Table 5.1 of that reference. We have also
numerically evaluated $\Phi$ for several HSD and found a very good agreement with
the values already obtained so we will not repeat them here.

As for the model dependent part of radiative corrections, we cannot compute it rigorously. Reference \cite{gk}, however,
proposes as a parametrization of this model dependence a modified weak coupling constant $G \equiv G (1+C)$, where $C\sim
0.0234$. This value of $C$ could give a noticeable contribution to the total decay rate. We will adopt this approach
in the present analysis.

\subsection{Experimental data on HSD}

The experimentally measured quantities \cite{part} in HSD are the total decay rate $R$, angular correlation coefficients
$\alpha_{e\nu}$, and angular spin-asymmetry coefficients $\alpha_e$, $\alpha_\nu$, $\alpha_B$, $A$, and $B$. An alternative
set of experimental data is constituted by the decay rates and measured $g_1/f_1$ ratios. This latter set, however, is not as
rich as the former and will not be used in the present analysis, unless noted otherwise. Currently there are five HSD which
have sufficient data to reliably extract the value of $V_{us}$. These processes are
$\Lambda \to p e^- \overline \nu_e$, $\Sigma^- \to n e^-\overline \nu_e$,
$\Xi^- \to \Lambda e^- \overline \nu_e$, $\Xi^-\to \Sigma^0 e^- \overline \nu_e$, and
$\Xi^0\to \Sigma^+ e^- \overline \nu_e$. Their available experimental information is displayed in Table
\ref{t:tab1}.

\begingroup
\squeezetable
\begin{table}
\caption{\label{t:tab1} Experimental data on five measured $|\Delta S| = 1$ HSD. The units of $R$ are $10^6 \, \textrm{s}^{-1}$.}
\begin{center}
\begin{tabular}{
l
r@{.}l@{\,$\pm$\,}r@{.}l r@{.}l@{\,$\pm$\,}r@{.}l
r@{.}l@{\,$\pm$\,}r@{.}l r@{.}l@{\,$\pm$\,}r@{.}l
r@{.}l@{\,$\pm$\,}r@{.}l
} \hline \hline
&
\multicolumn{4}{c}{$\Lambda \to p e^- \overline \nu_e$} &
\multicolumn{4}{c}{$\Sigma^- \to n e^-\overline \nu_e$} &
\multicolumn{4}{c}{$\Xi^- \to \Lambda e^- \overline \nu_e$} &
\multicolumn{4}{c}{$\Xi^-\to \Sigma^0 e^- \overline \nu_e$} &
\multicolumn{4}{c}{$\Xi^0\to \Sigma^+ e^- \overline \nu_e$} \\
\hline
$R$ &
3 & 161 & 0 & 058 & 6 & 88 & 0 & 24 & 3 & 44 & 0 & 19 &
0 & 53 & 0 & 10 & 0 & 93 & 0 & 14 \\
$\alpha_{e\nu}$ &
$-$0 & 019 & 0 & 013 & 0 & 347 & 0 & 024 & 0 & 53 & 0 & 10 &
\multicolumn{4}{c}{} & \multicolumn{4}{c}{} \\
$\alpha_e$ &
0 & 125 & 0 & 066 & $-$0 & 519 & 0 & 104 & \multicolumn{4}{c}{} &
\multicolumn{4}{c}{} & \multicolumn{4}{c}{} \\
$\alpha_\nu$ &
0 & 821 & 0 & 060 & $-$0 & 230 & 0 & 061 & \multicolumn{4}{c}{} &
\multicolumn{4}{c}{} & \multicolumn{4}{c}{} \\
$\alpha_B$ &
$-$0 & 508 & 0 & 065 & 0 & 509 & 0 & 102 & \multicolumn{4}{c}{} &
\multicolumn{4}{c}{} & \multicolumn{4}{c}{} \\
$A$ &
\multicolumn{4}{c}{} & \multicolumn{4}{c}{} & 0 & 62 & 0 & 10 &
\multicolumn{4}{c}{} & \multicolumn{4}{c}{} \\
$g_1/f_1$ &
0 & 718 & 0 & 015 & $-$0 & 340 &
0 & 017 & 0 & 25 & 0 & 05 &
1 & 287 & 0 & 158 & 1 & 32 & 0 & 22 \\ \hline \hline
\end{tabular}
\end{center}
\end{table}
\endgroup

\section{\label{sec:ncexp}HSD form factors in the $1/N_c$ expansion of QCD}

In the past when data were not very precise, fits to HSD were made under  the assumption of exact $SU(3)$ symmetry in order
to extract $V_{us}$. Currently, the experiments are precise enough to the extent that this assumption no longer provides a
reliably fit. Therefore, the determination of $V_{us}$ from HSD requires an understanding of the $SU(3)$ symmetry breaking effects in the
weak form factors. We devote this section to evaluate these effects within the framework of the $1/N_c$ expansion of QCD. The
form factors are analyzed in a combined expansion in $1/N_c$ and $SU(3)$ symmetry breaking following the lines of
Refs.~\cite{dash95,dai,rfm98}. Before doing so we first review some necessary large-$N_c$ formalism.

For large $N_c$, the lowest-lying baryons are given by the completely symmetric spin-flavor representation of $N_c$ quarks.
Under $SU(2) \times SU(N_F)$, this $SU(2N_F)$ representation decomposes into a tower of baryon flavor representations with
spins $J=\frac12, \frac32,\ldots,\frac{N_c}{2}$. For two flavors of light quarks the baryon tower consists of (spin,isospin)
representations with $I=J$, whereas for three flavors the baryon flavor representations become much more complex
\cite{dash94,dash95}.

In order to simplify the analysis, it is much better to concentrate on the baryon
operators, rather than on the states, because the former have a simple expansion in $1/N_c$ for arbitrary $N_c$. In this
context, the general form of the $1/N_c$ expansion of a QCD $m$-body quark operator acting on a single baryon state can be
written as \cite{dash94,dash95}
\begin{equation}
{\mathcal O}_{\textrm{QCD}}^{m\textrm{-body}} = N_c^m \sum_{n=0}^{N_c} c_n \frac{1}{N_c} {\mathcal O}_n, \label{eq:mbody}
\end{equation}
where $c_n$ are unknown coefficients which have power series expansions in $1/N_c$ beginning at order unity. The sum
in Eq.~(\ref{eq:mbody}) is over all possible independent $n$-body operators
${\mathcal O}_n$, $0\leq n \leq N_c$, with the same spin and flavor quantum numbers as ${\mathcal O}_{\textrm{QCD}}$. The use
of operator identities \cite{dash95} reduces the operator basis to independent operators. The large-$N_c$ spin-flavor
symmetry for baryons is generated by the baryon spin, flavor and spin-flavor operators $J^i$, $T^a$, and $G^{ia}$ which
can be written for large but finite $N_c$ as one-body quark operators acting on the $N_c$-quark baryon states as
\begin{subequations}
\label{eq:gen}
\begin{eqnarray}
J^i & = & q^\dagger \left(\frac{\sigma^i}{2} \otimes \openone \right) q \qquad (1, 1), \\
T^a & = & q^\dagger \left( \openone \otimes \frac{\lambda^a}{2} \right) q \qquad (0, 8), \\
G^{ia} & = & q^\dagger \left( \frac{\sigma^i}{2} \otimes \frac{\lambda^a}{2} \right) q \qquad (1, 8).
\end{eqnarray}
\end{subequations}
The transformation properties of these generators under $SU(2)\times SU(3)$ are given explicitly in Eq.~(\ref{eq:gen})
as $(j,d)$, where $j$ is the spin and $d$ is the dimension of the $SU(3)$ flavor representation.

In this paper we analyze the $1/N_c$ expansions of the QCD baryon vector and axial vector currents whose matrix elements between
$SU(6)$ symmetric states give the HSD form factors. The detailed analysis has already been done \cite{dai,rfm98},
so we will limit ourselves to only state the answer here.

\subsection{Vector form factor $f_1$}

At $q^2=0$ the hyperon matrix elements for the vector current are given by the matrix elements of the associated charge or
$SU(3)$ generator. The flavor octet baryon charge is denoted by \cite{rfm98}
\begin{equation}
V^{0a} = \left\langle B_2 \left|\left(\overline{q} \gamma^0 \frac{\lambda^a}{2} q \right)_{\textrm{QCD}} \right| B_1
\right\rangle
\end{equation}
and its matrix elements between $SU(6)$ symmetric states yield the value of $f_1$. $V^{0a}$ is spin-0 and a flavor
octet so that it transforms as (0,8) under $SU(2)\times SU(3)$. The $1/N_c$ expansion for the baryon vector current in the
limit of exact $SU(3)$ symmetry has the form
\begin{equation}
V^{0a} = \sum_{n=1}^{N_c} a_n \frac{1}{N_c^{n-1}} {\mathcal O}_n^a, \label{eq:vv0}
\end{equation}
where the allowed one- and two-body operators are ${\mathcal O}_1^a=T^a$ and ${\mathcal O}_2^a=\{J^i,G^{ia}\}$. Higher order
operators are obtained from the former as ${\mathcal O}_{n+2}^a =\{J^2,{\mathcal O}_n^a\}$.
The fact that at $q^2=0$ the baryon vector current $V^{0a}$ is
the generator of $SU(3)$ symmetry transformations imposes $a_1=1$ and $a_n=0$ for $n\geq 2$ in expansion (\ref{eq:vv0}).
Therefore, in this limit one has \cite{rfm98}
\begin{equation}
V^{0a} = T^a,
\end{equation}
whose matrix elements are denoted hereafter as $f_1^{SU(3)}$.

Flavor $SU(3)$ symmetry breaking in QCD is due to the light quark masses and transforms as a flavor
octet. The $SU(3)$ symmetry breaking correction to $V^{0a}$ was computed to second order in symmetry breaking in Ref.~\cite{rfm98},
as stated by the Ademollo-Gatto theorem \cite{ag}. The final expression
for the $1/N_c$ expansion of $V^{0a}$ can be cast into
\begin{equation}
V^{0a} = (1+v_1) T^a + v_2 \{T^a,N_s\} + v_3 \{T^a,-I^2+J_s^2\}, \label{eq:v0a}
\end{equation}
where $v_i$ are parameters to be determined. Besides,
$N_s$ is the number of strange quarks, $I$ is the isospin, and $J_s$ is the strange quark spin. The matrix
elements of the operators involved in the expansion (\ref{eq:v0a}) can be found in Ref.~\cite{rfm98} as well.

\subsection{Axial vector form factor $g_1$}

The $1/N_c$ expansion for the baryon axial-vector current $A^{ia}$ was first discussed in Refs.~\cite{dash95,dai}. We
will use a simplified version of their results here. For the $|\Delta S|=1$ sector of HSD, $A^{ia}$ can be written as
\begin{equation}
\frac12 A^{ia} = a^\prime G^{ia} + b^\prime J^iT^a + c_3 \{G^{ia},N_s\} + c_4 \{T^a,J_s^i\}. \label{eq:aia}
\end{equation}
Previous works \cite{dai,rfm98} included an extra term in expansion (\ref{eq:aia}) to account for strangeness-zero decays.
Adding this term avoided the mixing between symmetry breaking effects and $1/N_c$ corrections in the symmetric couplings $D$,
$F$, and ${\mathcal C}$. In our case such a term in not necessary, so we have removed it and kept only those terms
which contribute to strangeness-changing processes. This results in redefinitions of the parameters $a$ and $b$ of these
references into $a^\prime$ and $b^\prime$, which absorb the terms $c_1$ and $c_2$, respectively, of the original
expansion. The couplings $D$ and $F$ have to be redefined accordingly. For $A^{ia}$ we are thus left with four parameters,
namely, $a^\prime$, $b^\prime$, $c_3$ and $c_4$.

\subsection{The form factors $f_2$ and $g_2$}

The contributions of $f_2$ and $g_2$ to the decay amplitudes are suppressed by the momentum transfer. In the symmetry limit
the hyperon masses are degenerate and then such contributions vanish. Thus, the first-order symmetry breaking corrections to $f_2$ and $g_2$
actually contribute to second order in the decay amplitude.

In the limit of exact $SU(3)$ flavor symmetry the form factor $f_2$ is described by two invariants, $m_1$ and $m_2$, which
can be determined from the anomalous magnetic moments of the
nucleons \cite{dai}. The magnetic moment is a spin-1 octet operator so it has a $1/N_c$ expansion identical in structure to the
baryon axial-vector current $A^{ia}$ \cite{dash95,dai}. Nevertheless, it has been shown that reasonable shifts from the $SU(3)$
predictions of $f_2$ have no perceptible effects upon $\chi^2$ or $g_1$ in a global fit to experimental data
\cite{rfm96,aug}. We therefore
follow these references and determine $f_2$ with the best fit values $m_1=2.87$ and $m_2=-0.77$ \cite{dai}.

As for the form factor $g_2$, it vanishes in the $SU(3)$ flavor symmetry limit, so it is proportional to
$SU(3)$ symmetry breaking at leading order. The $1/N_c$ expansion for this form factor is given in detail in Ref.~\cite{rfm98}, where an
attempt was made in order to extract some quantitative information about it. However, it was concluded that the
experimental data are not precise enough for the extraction of the small $g_2$-dependence of the decay amplitudes. We
take the value $g_2=0$ in our analysis accordingly.

\section{\label{sec:rates}Fits to experimental data: decay rates and angular coefficients}

At this point we are now in a position to perform detailed comparisons with the experimental data of Table \ref{t:tab1}
through a number of fits. The experimental data which are used are the decay rates and the spin and angular correlation
coefficients of the five HSD listed. The value of the ratio $g_1/f_1$ is not used since it is determined from other
quantities and is not an independent measurement. For the processes $\Xi^-\to \Sigma^0 e^- \overline \nu_e$ and
$\Xi^0\to \Sigma^+ e^- \overline \nu_e$, however, we have no other choice but to use $g_1/f_1$ because no information on the
angular coefficients is available yet. The theoretical expressions for the total decay rates and angular
coefficients are organized in several tables in Appendix A. In the analysis we also take into account
both model-independent and model-dependent radiative corrections and the $q^2$-dependence of the leading form factors,
as stated in Sec.~\ref{sec:hsd}.

The parameters to be fitted are those arising out of the $1/N_c$ expansions of the baryon operators whose matrix
elements between $SU(6)$ symmetric states give the values of the couplings, namely,
$v_{1-3}$ for $f_1$ [introduced in Eq.~(\ref{eq:v0a})] and $a^\prime$, $b^\prime$, $c_{3-4}$ for $g_1$ [introduced in Eq.~(\ref{eq:aia})]. We
use the values of $f_2$ and $g_2$ in the limit of exact $SU(3)$ flavor symmetry. An additional input is the value
of $V_{us}$, Eq.~(\ref{eq:vuskl3}), which is mainly the one from $K_{e3}$ decays. We also extract information
on $V_{us}$ by fitting it as well. Hereafter, the quoted errors of the best fit parameters will be from the $\chi^2$ fit
only, and will not include any theoretical uncertainties.

\subsection{Exact $SU(3)$ symmetry}\label{sec:ft1}

As a starting point we can perform a rough $SU(3)$ symmetric fit which involves only the parameters $a^\prime$ and
$b^\prime$ for $g_1$.
Our aim is not quite to test the $1/N_c$ predictions but rather to explore the quality of the data of Table \ref{t:tab1}.
The results are displayed in the second column of Table \ref{t:tra}, labeled as Fit 1(a). We can immediately notice some
interesting results. As expected, the leading parameter $a^\prime$ is order unity and $b^\prime$ is order $1/N_c$, in good agreement with
previous works \cite{dai,rfm98}. In this case $\chi^2=38.63$ for 15 degrees of freedom. From the $\chi^2$ point of view,
the fit is very poor. The large value of $\chi^2$ is built up mainly by
$\alpha_e$ $(\Delta \chi^2=2.83)$ and $\alpha_\nu$ $(\Delta \chi^2=6.89)$ in $\Lambda \to p e^- \overline \nu_e$,
$R$ $(\Delta \chi^2=3.73)$, $\alpha_\nu$ $(\Delta \chi^2=4.04)$, and $\alpha_B$ $(\Delta \chi^2=2.32)$ in
$\Sigma^- \to n e^-\overline \nu_e$, and finally $R$ $(\Delta \chi^2=11.46)$ and $A$ $(\Delta \chi^2=2.05)$ in
$\Xi^- \to \Lambda e^- \overline \nu_e$.

We proceed to perform a similar fit but now with $V_{us}$ as a free parameter, along with $a^\prime$ and
$b^\prime$. This fit
is equivalent to the one recently performed in Ref.~\cite{cabb}, except that in this reference the decay rates and
$g_1/f_1$ ratios were used instead.
The results of our fit correspond to the third column of Table \ref{t:tra}, labeled as Fit 1(b).
There is a slight modification in the value of $a^\prime$ compared with the previous fit whereas $b^\prime$
remains practically unchanged. The fit yields $V_{us} =0.2238 \pm 0.0019$, which is lower than the one of Ref.~\cite{cabb}.
This time, $\chi^2=34.38$ for 14 degrees of freedom. Though $\chi^2$ is reduced by around 4, this is not much and the
fit is again far from being satisfactory. The lowering of $\chi^2$ comes mainly from $R$ in
$\Sigma^- \to n e^-\overline \nu_e$ and $\Xi^- \to \Lambda e^- \overline \nu_e$, whose contributions are reduced by
almost two and three, respectively. Still, $\alpha_\nu$ in $\Lambda \to p e^- \overline \nu_e$ and $R$ in
$\Xi^- \to \Lambda e^- \overline \nu_e$ show worrisome deviations from the theoretical predictions.

We close this section by pointing out that the high $\chi^2$ of these two fits is a clear evidence of $SU(3)$
symmetry breaking. We now proceed to analyze such effects by incorporating first- and second-order symmetry breaking into
the axial-vector and vector form factors $g_1$ and $f_1$, respectively.

\subsection{Symmetry breaking in $g_1$}\label{sec:ft2}

To appreciate the effects of the departure from the exact $SU(3)$ flavor symmetry, we incorporate first-order symmetry breaking
in $g_1$ through the parameters $a^\prime$, $b^\prime$ and $c_{3-4}$, while still keeping
$f_1$, $f_2$, and $g_2$ at their $SU(3)$ symmetric values.
Fitting these parameters leads to the results displayed as Fit 2(a) of Table \ref{t:tra}, with $\chi^2=24.41$ for 13
degrees of freedom. The highest contributions to $\chi^2$ come now from
$\alpha_e$ $(\Delta \chi^2=2.70)$ and $\alpha_\nu$ $(\Delta \chi^2=6.80)$ in $\Lambda \to p e^- \overline \nu_e$ and
$R$ $(\Delta \chi^2=2.55)$, $\alpha_\nu$ $(\Delta \chi^2=3.94)$, and $\alpha_B$ $(\Delta \chi^2=2.54)$ in
$\Sigma^- \to n e^-\overline \nu_e$. Except for the remarkable
improvement in the predictions of the observables in
$\Xi^- \to \Lambda e^- \overline \nu_e$, whose combined contribution to $\chi^2$ in this case amounts to less that 1.5,
we observe only slight reductions in the contributions to $\chi^2$ of the remaining observables, compared with Fit 1(a).
As for the fitted parameters, again the leading parameter
$a^\prime$ is order unity and $b^\prime$ is order $1/N_c$. The small effects due to symmetry breaking can be seen mainly in the new value
of $b^\prime$ compared to the $SU(3)$ symmetric fit, and in the parameters $c_{3-4}$, which are small or even smaller
than expected from first-order symmetry breaking (our rough measure of symmetry breaking is $\epsilon\sim 30\%$) and factors of $1/N_c$.
The values of the best fit parameters are consistent with previous works \cite{dai,rfm98}.

In a similar fashion, we can attempt to extract the value of $V_{us}$ in this context. The results are displayed
in the column labeled as Fit 2(b) of Table \ref{t:tra}. The fitted parameters change a little and
$V_{us}=0.2230 \pm 0.0019$ with $\chi^2=21.79$ for 12 degrees of freedom. The contributions to $\chi^2$
come from the very same observables as in the previous fit, with some minor changes in the observables other than the usual
ones which systematically have the highest contributions to $\chi^2$.
Regardless of the still high $\chi^2$, we can
observe that incorporating first-order symmetry breaking corrections into $g_1$ lowers the predicted value of $V_{us}$
compared to the case with no symmetry breaking at all, Fit 1(b). This fact indeed is crucial to reinforce
our initial argument that exact $SU(3)$ no longer provides an acceptable fit.

Let us now find out how the inclusion of second-order symmetry breaking into $f_1$ impacts on the various
observables, before drawing any conclusions.

\subsection{Symmetry breaking in both $f_1$ and $g_1$}\label{sec:ft3}

In this section we incorporate second-order symmetry breaking into the vector form factor $f_1$, so that it is no longer fixed
at its $SU(3)$ symmetric value $f_1^{SU(3)}$. We expect these effects to be second-order in symmetry breaking
(roughly $\epsilon^2\sim 9\%$) according to the Ademollo-Gatto theorem. For this fit, then, the parameters
$v_{1-3}$ of $f_1$ enter into play, simultaneously with $a^\prime$, $b^\prime$, and $c_{3-4}$ of $g_1$ while $f_2$ and $g_2$
remain fixed by exact $SU(3)$ symmetry. The best fit parameters are displayed as Fit 3(a) of Table \ref{t:tra}, with
$\chi^2=17.85$ for 10
degrees of freedom. When $V_{us}$ is also allowed to be a free parameter, we obtain the results displayed in the last
column of that table, labeled as Fit 3(b). The fit yields $V_{us}=0.2199 \pm 0.0026$. In both cases, the best fit parameters are as expected from
the $1/N_c$ expansion predictions. As a matter of fact, hereafter we will loosely refer to Fit 3(b) as the final fit.


\begingroup
\squeezetable
\begin{table}
\caption{\label{t:tra}Best fitted parameters for the vector and axial-vector form factors. The rates and asymmetry coefficients were used.}
\begin{center}
\begin{tabular}{
l
r@{\,$\pm$\,}l
r@{\,$\pm$\,}l
r@{\,$\pm$\,}l
r@{\,$\pm$\,}l
r@{\,$\pm$\,}l
r@{\,$\pm$\,}l
} \hline\hline
&
\multicolumn{4}{c}{Fit 1} &
\multicolumn{4}{c}{Fit 2} &
\multicolumn{4}{c}{Fit 3} \\
 & \multicolumn{2}{c}{(a)} & \multicolumn{2}{c}{(b)} & \multicolumn{2}{c}{(a)} &
\multicolumn{2}{c}{(b)} & \multicolumn{2}{c}{(a)} & \multicolumn{2}{c}{(b)} \\ \hline
$V_{us}$ & \multicolumn{2}{c}{Fixed} & 0.2238 & 0.0019 & \multicolumn{2}{c}{Fixed} &  0.2230 & 0.0019 &  \multicolumn{2}{c}{Fixed} &  0.2199 & 0.0026 \\
$v_1$ & \multicolumn{4}{c}{} & \multicolumn{4}{c}{} &    0.00 & 0.03   &   0.00 & 0.04 \\
$v_2$ & \multicolumn{4}{c}{} & \multicolumn{4}{c}{} &    0.02 & 0.03   &   0.02 & 0.03 \\
$v_3$ & \multicolumn{4}{c}{} & \multicolumn{4}{c}{} & $-0.01$ & 0.01   &$-0.01$ & 0.01 \\
$a^\prime$ & 0.80 & 0.01 &    0.78 & 0.01 &   0.71 & 0.03 &   0.70 & 0.03 &   0.72 & 0.03   & 0.72 & 0.03 \\
$b^\prime$ & $-0.07$ & 0.01 & $-0.07$ & 0.01 &   $-0.08$ & 0.01 &   $-0.08$ & 0.01 &  $-0.08$ & 0.01   & $-0.08$ & 0.01 \\
$c_3$ & \multicolumn{2}{c}{} & \multicolumn{2}{c}{} &  0.03 & 0.02 &   0.03 & 0.02 &   0.03 & 0.02 &  0.03 & 0.02 \\
$c_4$ & \multicolumn{2}{c}{} & \multicolumn{2}{c}{} &  0.06 & 0.02 &   0.06 & 0.02 &   0.05 & 0.02 &  0.05 & 0.02 \\
\hline
$\chi^2/\textrm{dof}$ & \multicolumn{2}{c}{38.63/15} & \multicolumn{2}{c}{34.38/14} & \multicolumn{2}{c}{24.41/13} &
\multicolumn{2}{c}{21.79/12} & \multicolumn{2}{c}{17.85/10} & \multicolumn{2}{c}{17.85/9} \\
\hline \hline
\end{tabular}
\end{center}
\end{table}
\endgroup

In Table \ref{t:pff} we display the predicted form factors corresponding to the final fit. These form factors yield
the predicted observables shown in Table \ref{t:pred}. Going through the latter table and comparing its entries with
the predictions produced by Fit 1(a), namely the $SU(3)$ fit, we can find some
improvements all over except in a well identified subset of data which carries most of the weight of the deviations from
the theoretical expectations. This subset is formed by the angular asymmetries $\alpha_e$, $\alpha_\nu$, and $\alpha_B$
of both processes $\Lambda \to p e^- \overline \nu_e$ and $\Sigma^- \to n e^-\overline \nu_e$,
which remain still too far from the current experimental data. Particularly,
there has been no noticeably change, in any fit performed, in either $\alpha_\nu$ or $\alpha_e$ of the former decay, despite
the important reduction of $\chi^2$ by more than half from the initial fit to the final one.


\begingroup
\squeezetable
\begin{table}
\caption{\label{t:pff}Predicted form factors. The quoted errors come from the fit only.}
\begin{center}
\begin{tabular}{
l
r@{.}l
r@{.}l
r@{.}l@{\,$\pm$\,}r@{.}l
r@{.}l@{\,$\pm$\,}r@{.}l
r@{.}l@{\,$\pm$\,}r@{.}l
r@{.}l@{\,$\pm$\,}r@{.}l
r@{.}l@{\,$\pm$\,}r@{.}l
r@{.}l@{\,$\pm$\,}r@{.}l
r@{.}l@{\,$\pm$\,}r@{.}l
} \hline \hline
&
\multicolumn{8}{c}{Fit 1(a)} & \multicolumn{4}{c}{Fit 1(b)} & \multicolumn{4}{c}{Fit 2(a)} & \multicolumn{4}{c}{Fit 2(b)} &
\multicolumn{8}{c}{Fit 3(b)} \\
Transition & \multicolumn{2}{c}{$f_1$} & \multicolumn{2}{c}{$f_2$} & \multicolumn{4}{c}{$g_1$} & \multicolumn{4}{c}{$g_1$} &
\multicolumn{4}{c}{$g_1$} & \multicolumn{4}{c}{$g_1$} & \multicolumn{4}{c}{$f_1$} & \multicolumn{4}{c}{$g_1$} \\ \hline
$\Lambda \to p$ &
$-$1 & 22 & $-$1 & 10 & $-$0 & 89 & 0 & 01 & $-$0 & 87 & 0 & 01 & $-$0 & 88 & 0 & 01 & $-$0 & 87 & 0 & 01 & $-$1 & 25 & 0 & 02 & $-$0 & 88 & 0 & 02 \\
$\Sigma^- \to n$ &
$-$1 & 00 & 1 & 02 & 0 & 34 & 0 & 01 & 0 & 33 & 0 & 01 & 0 & 35 & 0 & 01 & 0 & 34 & 0 & 01 & $-$1 & 04 & 0 & 02 & 0 & 34 & 0 & 01 \\
$\Xi^- \to \Lambda$ &
1 & 22 & $-$0 & 07 & 0 & 24 & 0 & 01 & 0 & 23 & 0 & 01 & 0 & 40 & 0 & 04 & 0 & 38 & 0 & 04 & 1 & 28 & 0 & 06 & 0 & 37 & 0 & 05 \\
$\Xi^-\to \Sigma^0$ &
0 & 71 & 1 & 31 & 0 & 89 & 0 & 01 & 0 & 87 & 0 & 01 & 0 & 92 & 0 & 05 & 0 & 91 & 0 & 05 & 0 & 75 & 0 & 04 & 0 & 93 & 0 & 06 \\
$\Xi^0\to \Sigma^+$ &
1 & 00 & 1 & 85 & 1 & 26 & 0 & 01 & 1 & 23 & 0 & 02 & 1 & 30 & 0 & 08 & 1 & 26 & 0 & 08 & 1 & 07 & 0 & 05 & 1 & 31 & 0 & 08 \\
\hline \hline
\end{tabular}
\end{center}
\end{table}
\endgroup


\begingroup
\squeezetable
\begin{table}
\caption{\label{t:pred}Theoretical predictions for five $|\Delta S| = 1$ hyperon
semileptonic decays and their contributions to the total $\chi^2$. The rates and angular coefficients were mainly used
in the fit. The units of $R$ are 10$^{6}$ s$^{-1}$.}
\begin{center}
\begin{tabular}{
l
r@{.}l@{\,\,\,\,}r@{.}l r@{.}l@{\,\,\,\,}r@{.}l r@{.}l@{\,\,\,\,}r@{.}l
r@{.}l@{\,\,\,\,}r@{.}l r@{.}l@{\,\,\,\,}r@{.}l r@{.}l@{\,\,\,\,}r@{.}l
r@{.}l@{\,\,\,\,}r@{.}l r@{.}l@{\,\,\,\,}r@{.}l r@{.}l@{\,\,\,\,}r@{.}l
r@{.}l@{\,\,\,\,}r@{.}l
} \hline \hline
&
\multicolumn{4}{c}{$\Lambda \to p e^- \overline \nu_e$} &
\multicolumn{4}{c}{$\Sigma^- \to n e^-\overline \nu_e$} &
\multicolumn{4}{c}{$\Xi^- \to \Lambda e^- \overline \nu_e$} &
\multicolumn{4}{c}{$\Xi^-\to \Sigma^0 e^- \overline \nu_e$} &
\multicolumn{4}{c}{$\Xi^0\to \Sigma^+ e^- \overline \nu_e$} \\ \hline
&
\multicolumn{2}{c}{Prediction} & \multicolumn{2}{c}{$\Delta \chi^2$} &
\multicolumn{2}{c}{Prediction} & \multicolumn{2}{c}{$\Delta \chi^2$} &
\multicolumn{2}{c}{Prediction} & \multicolumn{2}{c}{$\Delta \chi^2$} &
\multicolumn{2}{c}{Prediction} & \multicolumn{2}{c}{$\Delta \chi^2$} &
\multicolumn{2}{c}{Prediction} & \multicolumn{2}{c}{$\Delta \chi^2$} \\ \hline
$R$ &
3 & 16 & 0 & 0 & 6 & 87 & 0 & 0 & 3 & 40 & 0 & 0 & 0 & 54 & 0 & 1 & 0 & 98 & 0 & 1 \\
$\alpha_{e\nu}$ &
$-$0 & 01 & 0 & 1 & 0 & 36 & 0 & 1 & 0 & 51 & 0 & 0 & \multicolumn{8}{c}{} \\
$\alpha_e$ &
0 & 03 & 2 & 3 & $-$0 & 62 & 0 & 9 & \multicolumn{12}{c}{} \\
$\alpha_\nu$ &
0 & 97 & 6 & 5 & $-$0 & 35 & 3 & 8 & \multicolumn{12}{c}{} \\
$\alpha_B$ &
$-$0 & 59 & 1 & 7 & 0 & 65 & 2 & 0 & \multicolumn{12}{c}{} \\
$A$ &
\multicolumn{8}{c}{} & 0 & 63 & 0 & 1 & \multicolumn{8}{c}{} \\
$g_1/f_1$ &
0 & 71 & \multicolumn{2}{c}{} & $-$0 & 33 & \multicolumn{2}{c}{} &
0 & 29 & \multicolumn{2}{c}{} & 1 & 23 & 0 & 1 & 1 & 23 & 0 & 2 \\ \hline \hline
\end{tabular}
\end{center}
\end{table}
\endgroup

\section{\label{sec:alt}Fits to experimental data: decay rates and $g_1/f_1$ ratios}

We can now attempt to make a comparison between theory and experiment in another way. This time we can perform
a global fit by using the decay rates and measured $g_1/f_1$ ratios, the latter also contained in Table \ref{t:tab1}. We
proceed as before, namely, we first
perform an $SU(3)$ fit, next we include first- and second-order symmetry breaking effects in the axial-vector and vector
form factors, respectively, along the lines of Secs.~\ref{sec:ft1}-\ref{sec:ft3}. The results are all displayed in Table
\ref{t:tg1f1} as Fits 4, 5, and 6. Hereafter, let us refer to Fit 6(b) --the one with symmetry breaking effects in $f_1$
and $g_1$ and $V_{us}$ as a free parameter-- as the alternative fit.


\begingroup
\squeezetable
\begin{table}
\caption{\label{t:tg1f1}Best fitted parameters for the vector and axial-vector form factors. The rates and $g_1/f_1$ ratios were used.}
\begin{center}
\begin{tabular}{
l
r@{\,$\pm$\,}l
r@{\,$\pm$\,}l
r@{\,$\pm$\,}l
r@{\,$\pm$\,}l
r@{\,$\pm$\,}l
r@{\,$\pm$\,}l
} \hline\hline
&
\multicolumn{4}{c}{Fit 4} &
\multicolumn{4}{c}{Fit 5} &
\multicolumn{4}{c}{Fit 6} \\
 & \multicolumn{2}{c}{(a)} & \multicolumn{2}{c}{(b)} & \multicolumn{2}{c}{(a)} &
\multicolumn{2}{c}{(b)} & \multicolumn{2}{c}{(a)} & \multicolumn{2}{c}{(b)} \\ \hline
$V_{us}$ & \multicolumn{2}{c}{Fixed} & 0.2230 & 0.0019 & \multicolumn{2}{c}{Fixed} &  0.2222 & 0.0019 &  \multicolumn{2}{c}{Fixed} &  0.2200 & 0.0026 \\
$v_1$ & \multicolumn{4}{c}{} & \multicolumn{4}{c}{} & $-0.02$ & 0.04   & $-0.02$ & 0.04 \\
$v_2$ & \multicolumn{4}{c}{} & \multicolumn{4}{c}{} &    0.03 & 0.03   &   0.03 & 0.03 \\
$v_3$ & \multicolumn{4}{c}{} & \multicolumn{4}{c}{} & $-0.01$ & 0.01   &$-0.01$ & 0.01 \\
$a^\prime$ & 0.81 & 0.01 &    0.80 & 0.01 &   0.73 & 0.03 &   0.73 & 0.03 &   0.75 & 0.03   & 0.75 & 0.03 \\
$b^\prime$ & $-0.08$ & 0.01 & $-0.08$ & 0.01 &   $-0.09$ & 0.01 &   $-0.08$ & 0.01 &  $-0.08$ & 0.01   & $-0.08$ & 0.01 \\
$c_3$ & \multicolumn{2}{c}{} & \multicolumn{2}{c}{} &  0.02 & 0.02 &   0.02 & 0.02 &   0.02 & 0.02 &  0.02 & 0.02 \\
$c_4$ & \multicolumn{2}{c}{} & \multicolumn{2}{c}{} &  0.06 & 0.02 &   0.05 & 0.02 &   0.04 & 0.02 &  0.04 & 0.02 \\
\hline
$\chi^2/\textrm{dof}$ & \multicolumn{2}{c}{16.50/8} & \multicolumn{2}{c}{13.91/7} & \multicolumn{2}{c}{5.27/6} &
\multicolumn{2}{c}{3.86/5} & \multicolumn{2}{c}{0.72/3} & \multicolumn{2}{c}{0.72/2} \\
\hline \hline
\end{tabular}
\end{center}
\end{table}
\endgroup

The parameters involved in the fits follow a similar behavior as the preceding ones,
so there is no need to reproduce here the predicted form factors. Instead, we proceed to display in Table \ref{t:tg1f1} the
predicted observables obtained within the alternative fit. Looking through Tables \ref{t:tg1f1} and \ref{t:predrg1},
we find a very good agreement between the final fit and the alternative one. We also observe that the value of $V_{us}$
is systematically reduced from the $SU(3)$ prediction by including symmetry breaking effects in the
form factors. The alternative fit yields $V_{us}=0.2200\pm 0.0026$, in good agreement with the final fit value.
Indeed, taking into account the low $\chi^2$ of the alternative fit, we might conclude that it is satisfactory. This
conclusion is misleading because fitting the rates and $g_1/f_1$ ratios hides
the deviations in the polarization data found in Sec.~\ref{sec:rates}. This interesting finding cannot be elucidated otherwise.


\begingroup
\squeezetable
\begin{table}
\caption{\label{t:predrg1}Theoretical predictions for five $|\Delta S| = 1$ hyperon
semileptonic decays and their contributions to the total $\chi^2$. The rates and $g_1/f_1$ ratios were
used in the fit. The units of $R$ are 10$^{6}$ s$^{-1}$.}
\begin{center}
\begin{tabular}{
l
r@{.}l@{\,\,\,\,}r@{.}l r@{.}l@{\,\,\,\,}r@{.}l r@{.}l@{\,\,\,\,}r@{.}l
r@{.}l@{\,\,\,\,}r@{.}l r@{.}l@{\,\,\,\,}r@{.}l r@{.}l@{\,\,\,\,}r@{.}l
r@{.}l@{\,\,\,\,}r@{.}l r@{.}l@{\,\,\,\,}r@{.}l r@{.}l@{\,\,\,\,}r@{.}l
r@{.}l@{\,\,\,\,}r@{.}l
} \hline \hline
&
\multicolumn{4}{c}{$\Lambda \to p e^- \overline \nu_e$} &
\multicolumn{4}{c}{$\Sigma^- \to n e^-\overline \nu_e$} &
\multicolumn{4}{c}{$\Xi^- \to \Lambda e^- \overline \nu_e$} &
\multicolumn{4}{c}{$\Xi^-\to \Sigma^0 e^- \overline \nu_e$} &
\multicolumn{4}{c}{$\Xi^0\to \Sigma^+ e^- \overline \nu_e$} \\ \hline
&
\multicolumn{2}{c}{Prediction} & \multicolumn{2}{c}{$\Delta \chi^2$} &
\multicolumn{2}{c}{Prediction} & \multicolumn{2}{c}{$\Delta \chi^2$} &
\multicolumn{2}{c}{Prediction} & \multicolumn{2}{c}{$\Delta \chi^2$} &
\multicolumn{2}{c}{Prediction} & \multicolumn{2}{c}{$\Delta \chi^2$} &
\multicolumn{2}{c}{Prediction} & \multicolumn{2}{c}{$\Delta \chi^2$} \\ \hline
$R$ &
3 & 16 & 0 & 0 & 6 & 85 & 0 & 0 & 3 & 40 & 0 & 0 & 0 & 55 & 0 & 0 & 0 & 99 & 0 & 2 \\
$\alpha_{e\nu}$ &
$-$0 & 02 & \multicolumn{2}{c}{} & 0 & 34 & \multicolumn{2}{c}{} & 0 & 55 &  \multicolumn{2}{c}{}& \multicolumn{8}{c}{} \\
$\alpha_e$ &
0 & 02 & \multicolumn{2}{c}{} & $-$0 & 63 & \multicolumn{2}{c}{} & \multicolumn{12}{c}{} \\
$\alpha_\nu$ &
0 & 98 & \multicolumn{2}{c}{} & $-$0 & 35 & \multicolumn{2}{c}{} & \multicolumn{12}{c}{} \\
$\alpha_B$ &
$-$0 & 59 & \multicolumn{2}{c}{} & 0 & 66 & \multicolumn{2}{c}{} & \multicolumn{12}{c}{} \\
$A$ &
\multicolumn{8}{c}{} & 0 & 66 & \multicolumn{2}{c}{} & \multicolumn{8}{c}{} \\
$g_1/f_1$ &
0 & 72 & 0 & 0 & $-$0 & 34 & 0 & 0 &
0 & 26 & 0 & 1 & 1 & 22 & 0 & 2 & 1 & 22 & 0 & 2 \\ \hline \hline
\end{tabular}
\end{center}
\end{table}
\endgroup

\section{\label{sec:mod}Comparing with models of $SU(3)$ symmetry breaking}

Various treatments of $SU(3)$ breaking effects in the HSD couplings have been explicitly computed in order to understand the deviations from
exact $SU(3)$. We of course do not pretend to be exhaustive, but a representative selection of such treatments can be
found in Refs.~\cite{and,don,kra,sch}. It is hard to assess the success of these models, mainly because their
approaches and/or assumptions are rather different. Some rely on quark models and others on chiral perturbation theory
or some variations of such methods. They explicitly provide $SU(3)$ breaking corrections to $f_1$, which are summarized in
Table \ref{t:sbp} as
the ratios $f_1/f_1^{SU(3)}$. We also include in this table the patterns obtained in the present paper with the final fit of
Sec.~\ref{sec:ft3} and the alternative fit of Sec.~\ref{sec:alt}, together with the one of Ref.~\cite{rfm98}, which was obtained
under the same assumptions of this work but by performing a combined fit of HSD (both $\Delta S=0$ and $|\Delta S|=1$ data)
and pionic decays of the decuplet baryons.
With this information, we can proceed to find out the trends of these models toward the determination of $V_{us}$. As for
$g_1$, Refs.~\cite{don,and} also provide its breaking pattern. In order to make a comparison on an equal footing of the
four models we find more convenient to leave $g_1$ as a free parameter. We now must resort to a model-independent determination
of $g_1$ which allows the extraction of symmetry breaking corrections from experiment in a way as general as possible. For
this purpose we can use the $1/N_c$ expansion and fit the parameters $a^\prime$, $b^\prime$, and $c_{3-4}$, or we can adapt
the approach of Ref.~\cite{gk}, which assumes that symmetry breaking comes from the eight component of an octet in the
strong-interaction Hamiltonian. In this scheme, $g_1$ can be parametrized in terms of seven quantities, namely, $\tilde{F}$,
$\tilde{D}$, $A_1$,
$B_1$, $C_1$, $D_1$ and $E_1$, the first two quantities corresponding to the exact symmetric limit \footnote{Actually,
$\tilde{F}$ and $\tilde{D}$ are related to the conventional reduced form factors $F$ and $D$ as $F=\tilde{F}/\sqrt{6}$ and
$D =-\sqrt{3/10} \tilde{F}$. We have put a tilde on them to avoid confusion.}. For the fits we include
the decay rates, the angular coefficients and the ratio $g_1/f_1$ of $\Xi^- \to \Sigma^0 e^- \overline{\nu}$,
leaving out the experimental
information on the decay $\Xi^0\to \Sigma^+ e^- \overline{\nu}$ because the ratio $f_1/f_1^{SU(3)}$ is not provided
by the models.

The values of $V_{us}$ extracted within these models are listed in the bottom row of Table \ref{t:sbp}. The fits in general
are stable but produce $\chi^2/\textrm{dof}$ higher than two. Among the models, only Refs.~\cite{don,sch} quote values of
$V_{us}$ and our predictions agree well with theirs.
Starting with the $V_{us}$ obtained in the frame of exact $SU(3)$ symmetry in Fit 1(b), namely,
$V_{us} = 0.2238 \pm 0.0019$, we can observe immediately by looking through Table \ref{t:sbp} that
a pattern of symmetry breaking such as $f_1/f_1^{SU(3)} < 1$ will systematically increase $V_{us}$ from
the former value, whereas the opposite trend occurs when $f_1/f_1^{SU(3)} > 1$. Let us now discuss the consequences
of these findings in our concluding section.


\begingroup
\squeezetable
\begin{table}
\caption{\label{t:sbp}Symmetry breaking pattern for $f_1$. The entries correspond to $f_1/f_1^{\rm SU(3)}$.}
\begin{center}
\begin{tabular}{lccccccc}
\hline \hline
Transition &
Fit 3(b) & Fit 6(b) & R.F.M \textit{et.\ al} \cite{rfm98} & Anderson and Luty \cite{and} & Donoghue \textit{et.\ al.} \cite{don} &
Krause \cite{kra} & Schlumpf \cite{sch} \\
\hline
$\Lambda \to p$ &
1.02$\pm$0.02 & 1.02$\pm$0.02 & 1.02$\pm$0.02 & 1.024 & 0.987 & 0.943 & 0.976 \\
$\Sigma^- \to n$ &
1.04$\pm$0.02 & 1.04$\pm$0.03 & 1.04$\pm$0.02 & 1.100 & 0.987 & 0.987 & 0.975 \\
$\Xi^- \to \Lambda$ &
1.04$\pm$0.04 & 1.04$\pm$0.04 & 1.10$\pm$0.04 & 1.059 & 0.987 & 0.957 & 0.976 \\
$\Xi^-\to \Sigma^0$ &
1.07$\pm$0.05 & 1.08$\pm$0.05 & 1.12$\pm$0.05 & 1.011 & 0.987 & 0.943 & 0.976 \\
$\Xi^0\to \Sigma^+$ &
1.07$\pm$0.05 & 1.08$\pm$0.05 & 1.12$\pm$0.05 & & & \\[1.0mm] \hline
$V_{us}$ & 0.2199$\pm$0.0026 & 0.2200$\pm$0.0026 & 0.2194$\pm$0.0023 & 0.2177$\pm$0.0019 & 0.2244$\pm$0.0019 & 0.2274$\pm$0.0019 &
0.2256$\pm$0.0019 \\[1.0mm] \hline \hline
\end{tabular}
\end{center}
\end{table}
\endgroup

\section{\label{sec:concl}Conclusions}

So far we can establish two interesting findings from our analysis. The first one concerns the issue that the assumption
of exact $SU(3)$ symmetry, as customarily used to compare theory and experiment in HSD, is questionable due to
the poor fits it produces. The second one is related to the fact that deviations from the exact $SU(3)$ limit, accounted for
in the form factors $f_1$ and $g_1$ are indeed important in order to reliably determine $V_{us}$ from HSD, which
can rival in precision with the one from $K_{e3}$ decays. The value currently admitted of $V_{us}$ for the latter decays
is given in Eq.~(\ref{eq:vuskl3}).
In our analysis, we have performed a series of fits under several assumptions in the context of exact and broken $SU(3)$ symmetry.
If we consider the limit of exact symmetry, the fit produces $V_{us} = 0.2238 \pm 0.0019$, which is higher than (\ref{eq:vuskl3}).
By including first-order symmetry breaking effects into the axial-vector form factors $g_1$, the fit now
yields $V_{us}=0.2230 \pm 0.0019$, which is still higher. However,
the main conclusion we can draw from the above analysis is that a reconciliation between these two determinations
can be obtained only through second-order breaking effects of a few percent in the leading
vector form factors $f_1$, which always increase their magnitudes over their exact $SU(3)$ symmetry predictions $f_1^{SU(3)}$,
as displayed in Table \ref{t:sbp}, second column, and also found in previous works \cite{aug,and,rfm98}. Therefore,
experimental data seems to favor this trend. From this, the value of $V_{us}$ we can extract from hyperon semileptonic
decays is
\begin{equation}
V_{us} = 0.2199 \pm 0.0026, \label{eq:dsh}
\end{equation}
which is comparable to (\ref{eq:vuskl3}) and indeed, agrees very well with the value of $V_{us}=0.2208 \pm 0.0034$
obtained very recently from hadronic $\tau$ decays \cite{pich}.

We consider pertinent to remark that the experimental information used in the fit was constituted mainly by the decay rates and
angular correlation and spin-asymmetry coefficients. Although we have performed similar fits
by using the rates and the $g_1/f_1$ ratios, we should point out that the total $\chi^2$ corresponding to them is
small when symmetry breaking effects into the leading form factors are included.
If these were the only pieces of data available to us, instead of the angular coefficients, we would be prompted to conclude
that the exact $SU(3)$ symmetry limit is in very good agreement with experiment. It is clear that using the angular coefficients,
instead of the $g_1/f_1$ ratios, not only avoids inconsistencies but provides a more sensitive test. From this point of
view we can conclude that, although HSD data are rather scarce, they are restrictive enough to make us look into the
exact symmetry limit assumption with more care.

\acknowledgments
The author wishes to express his gratitude to J.~Engelfried, M.~Kirchbach, and A.~Morelos for their useful comments on the
manuscript. He also is grateful to Consejo Nacional de Ciencia y Tecnolog{\'\i}a and Fondo de Apoyo a la Investigaci\'on
(Universidad Aut\'onoma de San Luis Potos{\'\i}), Mexico, for partial support.

\appendix

\section{\label{sec:appa}Numerical formulas for HSD integrated observables}

In this Appendix we provide updated numerical formulas for the uncorrected transition rates and angular coefficients of the
five
$|\Delta S|=1$ HSD dealt with in the present paper, namely, $R^0$, and $\alpha_{e\nu}^0$, $\alpha_\nu^0$, $\alpha_e^0$,
$\alpha_B^0$, and $A^0$. The theoretical expressions used for these integrated observables
are those computed in Ref.~\cite{gk}, where one can find further details about the kinematical region of integration. The
inputs for the numerical evaluation are the value of the weak coupling constant $G$ and the experimental
values of the hyperon masses, which are all
found in Ref.~\cite{part}. The slope parameters of the leading form factors are defined in Eq.~(\ref{eq:slop}). Neither
radiative corrections nor $V_{us}$ are included at all.

The total decay rate $R^0$, being quadratic in the form factors, can be written in the most general form as
\begin{equation}
R^0 = \sum_{i\leq j=1}^6 a_{ij}^R f_i f_j + \sum_{i\leq j=1}^6 b_{ij}^R (f_i \lambda_{f_j} + f_j \lambda_{f_i}),
\label{eq:rnum}
\end{equation}
where dipole parametrizations similar to Eq.~(\ref{eq:slop}) have been assumed for all form factors, which introduce
in total six slope parameters $\lambda_{f_i}$. For the sake of shortening Eq.~(\ref{eq:rnum}), we have momentarily redefined
$g_1=f_4$, $g_2 =f_5$, $g_3=f_6$, $\lambda_{g_1}=\lambda_{f_4}$,
$\lambda_{g_2}=\lambda_{f_5}$, and $\lambda_{g_3}=\lambda_{f_6}$. Notice that the restriction
$i\leq j$ reduces each sum in Eq.~(\ref{eq:rnum}) to 21 terms.
Similar expressions to Eq.~(\ref{eq:rnum}) also hold for the products $R^0 \alpha^0$, where $\alpha^0$ is any of the
angular coefficients defined above. Once $R^0$ and $R^0 \alpha^0$ are determined, $\alpha^0$ is obtained straightforwardly.

The integrated observables have been organized in Tables \ref{t:t5}-\ref{t:t9}.
Although we have completely computed all 42 terms involved in Eq.~(\ref{eq:rnum}), we have not listed
neither the contributions of $f_3$ and $g_3$ nor the ones from the slope parameters $\lambda_{f_2}$, $\lambda_{g_2}$, $\lambda_{f_3}$,
and $\lambda_{g_3}$. The entries have been truncated to four decimal places so that we have also omitted
contributions lower than $10^{-5}$, which we consider small compared to the leading ones.

For a particular decay, the coefficients $a_{ij}^R$ and $b_{ij}^R$ of $R^0$ in Eq.~(\ref{eq:rnum}) can be easily read off from the second
column in each table. Similarly, in the third column one can read off the coefficients $a_{ij}^{R\alpha_{e\nu}}$ and
$b_{ij}^{R\alpha_{e\nu}}$ of the product $R^0\alpha_{e\nu}^0$, and so on. These numerical
formulas are the ones used in the fits to experimental data performed in the present paper.

\begingroup
\squeezetable
\begin{table}
\caption{\label{t:t5}Numerical formulas for some integrated observables of $\Lambda \to p e^- \overline{\nu}$ decay. The
units of $R^0$ are $10^6 \, \textrm{s}^{-1}$.}
\begin{center}
\begin{tabular}{
l
r@{.}l
r@{.}l
r@{.}l
r@{.}l
r@{.}l
r@{.}l
} \hline\hline
&
\multicolumn{2}{c}{$R^0$} &
\multicolumn{2}{c}{$R^0\alpha_{e\nu}^0$} &
\multicolumn{2}{c}{$R^0\alpha_\nu^0$} &
\multicolumn{2}{c}{$R^0\alpha_e^0$} &
\multicolumn{2}{c}{$R^0\alpha_B^0$} &
\multicolumn{2}{c}{$R^0A^0$} \\ \hline
$f_1f_1$ & 15 & 2774 & 12 & 5169 & 0 & 9291 & $-$0 & 9290 & \multicolumn{2}{c}{} & \multicolumn{2}{c}{} \\
$f_2f_2$ & 0 & 2200 & $-$0 & 1553 & 0 & 0860 & $-$0 & 0860 & \multicolumn{2}{c}{} & \multicolumn{2}{c}{} \\
$g_1g_1$ & 45 & 4432 & $-$22 & 4798 & 31 & 0949 & $-$31 & 0937 & \multicolumn{2}{c}{} & $-$0 & 0008 \\
$g_2g_2$ & 0 & 6558 & $-$0 & 8591 & 0 & 5217 & $-$0 & 5217 & \multicolumn{2}{c}{} & \multicolumn{2}{c}{} \\
$f_1f_2$ & 0 & 3580 & $-$0 & 1982 & 1 & 7571 & $-$1 & 7570 & \multicolumn{2}{c}{} & 0 & 0001 \\
$g_1g_2$ & $-$9 & 6247 & 11 & 1472 & $-$8 & 2254 & 8 & 2252 & \multicolumn{2}{c}{} & 0 & 0002 \\
$f_1g_1$ & \multicolumn{2}{c}{} & $-$0 & 0002 & 28 & 6966 & 28 & 6951 & $-$35 & 7929 & 41 & 9279 \\
$f_1g_2$ & \multicolumn{2}{c}{} & \multicolumn{2}{c}{} & $-$1 & 3993 & $-$1 & 3991 & 1 & 6632 & $-$2 & 8165 \\
$f_2g_1$ & \multicolumn{2}{c}{} & $-$0 & 0004 & $-$1 & 3992 & $-$1 & 3995 & 3 & 8473 & 2 & 8167 \\
$f_2g_2$ & \multicolumn{2}{c}{} & 0 & 0001 & 0 & 2681 & 0 & 2681 & $-$0 & 6256 & $-$0 & 4911 \\
$f_1 \lambda_{f_1}$ & 0 & 2211 & $-$0 & 0570 & 0 & 0201 & $-$0 & 0201 & \multicolumn{2}{c}{} & \multicolumn{2}{c}{} \\
$g_1 \lambda_{g_1}$ & 1 & 0926 & $-$1 & 4647 & 0 & 8915 & $-$0 & 8915 & \multicolumn{2}{c}{} & \multicolumn{2}{c}{} \\
$f_1 \lambda_{g_1} + g_1 \lambda_{f_1}$ & \multicolumn{2}{c}{} & \multicolumn{2}{c}{} & 0 & 2011 & 0 & 2011 & $-$0 & 2921 & 0 & 3683 \\
\hline \hline
\end{tabular}
\end{center}
\end{table}
\endgroup

\begingroup
\squeezetable
\begin{table}
\caption{\label{t:t6}Numerical formulas for some integrated observables of $\Sigma^- \to n e^- \overline{\nu}$ decay. The
units of $R^0$ are $10^6 \, \textrm{s}^{-1}$.}
\begin{center}
\begin{tabular}{
l
r@{.}l
r@{.}l
r@{.}l
r@{.}l
r@{.}l
r@{.}l
} \hline\hline
&
\multicolumn{2}{c}{$R^0$} &
\multicolumn{2}{c}{$R^0\alpha_{e\nu}^0$} &
\multicolumn{2}{c}{$R^0\alpha_\nu^0$} &
\multicolumn{2}{c}{$R^0\alpha_e^0$} &
\multicolumn{2}{c}{$R^0\alpha_B^0$} &
\multicolumn{2}{c}{$R^0A^0$} \\ \hline
$f_1f_1$ & 90 & 5903 & 67 & 3788 & 7 & 8358 & $-$7 & 8356 & \multicolumn{2}{c}{} & 0 & 0002 \\
$f_2f_2$ & 2 & 3860 & $-$1 & 8647 & 0 & 9864 & $-$0 & 9864 & \multicolumn{2}{c}{} & \multicolumn{2}{c}{} \\
$g_1g_1$ & 267 & 2903 & $-$147 & 6648 & 184 & 5359 & $-$184 & 5326 & \multicolumn{2}{c}{} & $-$0 & 0024 \\
$g_2g_2$ & 7 & 0659 & $-$9 & 6050 & 5 & 6662 & $-$5 & 6662 & \multicolumn{2}{c}{} & \multicolumn{2}{c}{} \\
$f_1f_2$ & 3 & 9980 & $-$2 & 5400 & 14 & 5368 & $-$14 & 5364 & \multicolumn{2}{c}{} & 0 & 0003 \\
$g_1g_2$ & $-$76 & 5914 & 92 & 9306 & $-$66 & 0522 & 66 & 0512 & \multicolumn{2}{c}{} & 0 & 0007 \\
$f_1g_1$ & \multicolumn{2}{c}{} & $-$0 & 0008 & 165 & 5089 & 165 & 5046 & $-$206 & 0087 & 250 & 6750 \\
$f_1g_2$ & \multicolumn{2}{c}{} & 0 & 0002 & $-$10 & 5392 & $-$10 & 5383 & 12 & 1688 & $-$23 & 2535 \\
$f_2g_1$ & \multicolumn{2}{c}{} & $-$0 & 0014 & $-$10 & 5388 & $-$10 & 5398 & 30 & 1444 & 23 & 2542 \\
$f_2g_2$ & \multicolumn{2}{c}{} & 0 & 0003 & 2 & 7993 & 2 & 7994 & $-$6 & 7007 & $-$5 & 4083 \\
$f_1 \lambda_{f_1}$ & 2 & 4092 & $-$0 & 8622 & 0 & 3096 & $-$0 & 3096 & \multicolumn{2}{c}{} & \multicolumn{2}{c}{} \\
$g_1 \lambda_{g_1}$ & 11 & 7689 & $-$16 & 3427 & 9 & 6693 & $-$9 & 6693 & \multicolumn{2}{c}{} & $-$0 & 0001 \\
$f_1 \lambda_{g_1} + g_1 \lambda_{f_1}$ & \multicolumn{2}{c}{} & \multicolumn{2}{c}{} & 2 & 0996 & 2 & 0995 & $-$3 & 0382 & 4 & 0562 \\
\hline \hline
\end{tabular}
\end{center}
\end{table}
\endgroup

\begingroup
\squeezetable
\begin{table}
\caption{\label{t:t7}Numerical formulas for some integrated observables of $\Xi^- \to \Lambda e^- \overline{\nu}$ decay. The
units of $R^0$ are $10^6 \, \textrm{s}^{-1}$.}
\begin{center}
\begin{tabular}{
l
r@{.}l
r@{.}l
r@{.}l
r@{.}l
r@{.}l
r@{.}l
} \hline\hline
&
\multicolumn{2}{c}{$R^0$} &
\multicolumn{2}{c}{$R^0\alpha_{e\nu}^0$} &
\multicolumn{2}{c}{$R^0\alpha_\nu^0$} &
\multicolumn{2}{c}{$R^0\alpha_e^0$} &
\multicolumn{2}{c}{$R^0\alpha_B^0$} &
\multicolumn{2}{c}{$R^0A^0$} \\ \hline
$f_1f_1$ & 32 & 1282 & 26 & 4627 & 1 & 9066 & $-$1 & 9065 & \multicolumn{2}{c}{} & 0 & 0001 \\
$f_2f_2$ & 0 & 4433 & $-$0 & 3108 & 0 & 1726 & $-$0 & 1726 & \multicolumn{2}{c}{} & \multicolumn{2}{c}{} \\
$g_1g_1$ & 95 & 6040 & $-$46 & 9613 & 65 & 3824 & $-$65 & 3805 & \multicolumn{2}{c}{} & $-$0 & 0013 \\
$g_2g_2$ & 1 & 3215 & $-$1 & 7273 & 1 & 0508 & $-$1 & 0508 & \multicolumn{2}{c}{} & \multicolumn{2}{c}{} \\
$f_1f_2$ & 0 & 7199 & $-$0 & 3951 & 3 & 6100 & $-$3 & 6099 & \multicolumn{2}{c}{} & 0 & 0001 \\
$g_1g_2$ & $-$19 & 8176 & 22 & 8864 & $-$16 & 9273 & 16 & 9268 & \multicolumn{2}{c}{} & 0 & 0003 \\
$f_1g_1$ & \multicolumn{2}{c}{} & $-$0 & 0003 & 60 & 4433 & 60 & 4410 & $-$75 & 3978 & 88 & 1274 \\
$f_1g_2$ & \multicolumn{2}{c}{} & \multicolumn{2}{c}{} & $-$2 & 8903 & $-$2 & 8900 & 3 & 4407 & $-$5 & 7868 \\
$f_2g_1$ & \multicolumn{2}{c}{} & $-$0 & 0006 & $-$2 & 8901 & $-$2 & 8906 & 7 & 9291 & 5 & 7871 \\
$f_2g_2$ & \multicolumn{2}{c}{} & 0 & 0001 & 0 & 5413 & 0 & 5413 & $-$1 & 2611 & $-$0 & 9883 \\
$f_1 \lambda_{f_1}$ & 0 & 4454 & $-$0 & 1121 & 0 & 0394 & $-$0 & 0394 & \multicolumn{2}{c}{} & \multicolumn{2}{c}{} \\
$g_1 \lambda_{g_1}$ & 2 & 2017 & $-$2 & 9451 & 1 & 7958 & $-$1 & 7957 & \multicolumn{2}{c}{} & \multicolumn{2}{c}{} \\
$f_1 \lambda_{g_1} + g_1 \lambda_{f_1}$ & \multicolumn{2}{c}{} & \multicolumn{2}{c}{} & 0 & 4060 & 0 & 4060 & $-$0 & 5899 & 0 & 7413 \\
\hline \hline
\end{tabular}
\end{center}
\end{table}
\endgroup

\begingroup
\squeezetable
\begin{table}
\caption{\label{t:t8}Numerical formulas for some integrated observables of $\Xi^- \to \Sigma^0 e^- \overline{\nu}$ decay. The
units of $R^0$ are $10^6 \, \textrm{s}^{-1}$.}
\begin{center}
\begin{tabular}{
l
r@{.}l
r@{.}l
r@{.}l
r@{.}l
r@{.}l
r@{.}l
} \hline\hline
&
\multicolumn{2}{c}{$R^0$} &
\multicolumn{2}{c}{$R^0\alpha_{e\nu}^0$} &
\multicolumn{2}{c}{$R^0\alpha_\nu^0$} &
\multicolumn{2}{c}{$R^0\alpha_e^0$} &
\multicolumn{2}{c}{$R^0\alpha_B^0$} &
\multicolumn{2}{c}{$R^0A^0$} \\ \hline
$f_1f_1$ & 3 & 3767 & 3 & 0211 & 0 & 1192 & $-$0 & 1192 & \multicolumn{2}{c}{} & \multicolumn{2}{c}{} \\
$f_2f_2$ & 0 & 0183 & $-$0 & 0114 & 0 & 0067 & $-$0 & 0067 & \multicolumn{2}{c}{} & \multicolumn{2}{c}{} \\
$g_1g_1$ & 10 & 0998 & $-$4 & 3598 & 6 & 8423 & $-$6 & 8418 & \multicolumn{2}{c}{} & $-$0 & 0004 \\
$g_2g_2$ & 0 & 0547 & $-$0 & 0687 & 0 & 0431 & $-$0 & 0431 & \multicolumn{2}{c}{} & \multicolumn{2}{c}{} \\
$f_1f_2$ & 0 & 0288 & $-$0 & 0134 & 0 & 2303 & $-$0 & 2303 & \multicolumn{2}{c}{} & \multicolumn{2}{c}{} \\
$g_1g_2$ & $-$1 & 3109 & 1 & 4382 & $-$1 & 1093 & 1 & 1093 & \multicolumn{2}{c}{} & \multicolumn{2}{c}{} \\
$f_1g_1$ & \multicolumn{2}{c}{} & $-$0 & 0001 & 6 & 5150 & 6 & 5144 & $-$8 & 1375 & 9 & 1718 \\
$f_1g_2$ & \multicolumn{2}{c}{} & \multicolumn{2}{c}{} & $-$0 & 2016 & $-$0 & 2015 & 0 & 2454 & $-$0 & 3694 \\
$f_2g_1$ & \multicolumn{2}{c}{} & $-$0 & 0001 & $-$0 & 2015 & $-$0 & 2016 & 0 & 5327 & 0 & 3694 \\
$f_2g_2$ & \multicolumn{2}{c}{} & \multicolumn{2}{c}{} & 0 & 0231 & 0 & 0231 & $-$0 & 0526 & $-$0 & 0401 \\
$f_1 \lambda_{f_1}$ & 0 & 0183 & $-$0 & 0028 & 0 & 0010 & $-$0 & 0010 & \multicolumn{2}{c}{} & \multicolumn{2}{c}{} \\
$g_1 \lambda_{g_1}$ & 0 & 0912 & $-$0 & 1173 & 0 & 0738 & $-$0 & 0738 & \multicolumn{2}{c}{} & \multicolumn{2}{c}{} \\
$f_1 \lambda_{g_1} + g_1 \lambda_{f_1}$ & \multicolumn{2}{c}{} & \multicolumn{2}{c}{} & 0 & 0173 & 0 & 0173 & $-$0 & 0253 & 0 & 0301 \\
\hline \hline
\end{tabular}
\end{center}
\end{table}
\endgroup

\begingroup
\squeezetable
\begin{table}
\caption{\label{t:t9}Numerical formulas for some integrated observables of $\Xi^0 \to \Sigma^+ e^- \overline{\nu}$ decay. The
units of $R^0$ are $10^6 \, \textrm{s}^{-1}$.}
\begin{center}
\begin{tabular}{
l
r@{.}l
r@{.}l
r@{.}l
r@{.}l
r@{.}l
r@{.}l
} \hline\hline
&
\multicolumn{2}{c}{$R^0$} &
\multicolumn{2}{c}{$R^0\alpha_{e\nu}^0$} &
\multicolumn{2}{c}{$R^0\alpha_\nu^0$} &
\multicolumn{2}{c}{$R^0\alpha_e^0$} &
\multicolumn{2}{c}{$R^0\alpha_B^0$} &
\multicolumn{2}{c}{$R^0A^0$} \\ \hline
$f_1f_1$ & 2 & 9853 & 2 & 6777 & 0 & 1031 & $-$0 & 1031 & \multicolumn{2}{c}{} & \multicolumn{2}{c}{} \\
$f_2f_2$ & 0 & 0155 & $-$0 & 0096 & 0 & 0057 & $-$0 & 0057 & \multicolumn{2}{c}{} & \multicolumn{2}{c}{} \\
$g_1g_1$ & 8 & 9302 & $-$3 & 8371 & 6 & 0480 & $-$6 & 0475 & \multicolumn{2}{c}{} & $-$0 & 0003 \\
$g_2g_2$ & 0 & 0464 & $-$0 & 0582 & 0 & 0366 & $-$0 & 0366 & \multicolumn{2}{c}{} & \multicolumn{2}{c}{} \\
$f_1f_2$ & 0 & 0244 & $-$0 & 0113 & 0 & 1993 & $-$0 & 1993 & \multicolumn{2}{c}{} & \multicolumn{2}{c}{} \\
$g_1g_2$ & $-$1 & 1358 & 1 & 2438 & $-$0 & 9608 & 0 & 9608 & \multicolumn{2}{c}{} & \multicolumn{2}{c}{} \\
$f_1g_1$ & \multicolumn{2}{c}{} & \multicolumn{2}{c}{} & 5 & 7644 & 5 & 7639 & $-$7 & 2001 & 8 & 1058 \\
$f_1g_2$ & \multicolumn{2}{c}{} & \multicolumn{2}{c}{} & $-$0 & 1749 & $-$0 & 1749 & 0 & 2131 & $-$0 & 3197 \\
$f_2g_1$ & \multicolumn{2}{c}{} & $-$0 & 0001 & $-$0 & 1749 & $-$0 & 1750 & 0 & 4618 & 0 & 3197 \\
$f_2g_2$ & \multicolumn{2}{c}{} & \multicolumn{2}{c}{} & 0 & 0197 & 0 & 0197 & $-$0 & 0446 & $-$0 & 0340 \\
$f_1 \lambda_{f_1}$ & 0 & 0155 & $-$0 & 0023 & 0 & 0008 & $-$0 & 0008 & \multicolumn{2}{c}{} & \multicolumn{2}{c}{} \\
$g_1 \lambda_{g_1}$ & 0 & 0774 & $-$0 & 0995 & 0 & 0627 & $-$0 & 0627 & \multicolumn{2}{c}{} & \multicolumn{2}{c}{} \\
$f_1 \lambda_{g_1} + g_1 \lambda_{f_1}$ & \multicolumn{2}{c}{} & \multicolumn{2}{c}{} & 0 & 0147 & 0 & 0147 & $-$0 & 0215 & 0 & 0255 \\
\hline \hline
\end{tabular}
\end{center}
\end{table}
\endgroup

\end{document}